\documentclass[aps,twocolumn]{revtex4}
\usepackage{epsf}
 
\begin{document}
      \title{Kondo impurity between superconducting and metallic
        reservoir:\\ the flow equation approach}

     \author{M.\ Zapalska and T.\ Doma\'nski}
\affiliation{
             Institute of Physics, 
	     M.\ Curie Sk\l odowska University, 
             20-031 Lublin, Poland} 
      \date{\today}

\begin{abstract}
It is well established that a correlated quantum impurity embedded in 
a metallic host can form the many-body Kondo state with itinerant electrons 
due to the effective antiferromagnetic coupling. Such effect is manifested 
spectroscopically by a narrow Abrikosov-Suhl peak appearing at the Fermi 
level below a characteristic temperature $T_{K}$. Recent experiments 
using  nanoscopic heterojunctions where the correlated quantum 
impurities (dots) are coupled to superconducting reservoirs revealed 
that the Kondo-type correlations are substantially weaker because: 
i) the single-particle states of superconductors are depleted around 
the Fermi level and ii) the on-dot pairing (proximity effect) competes 
with the spin ordering. Within the Anderson impurity scenario we study 
here influence of such induced on-dot paring on the exchange interaction 
adopting the continuous unitary transformation, which goes beyond 
the perturbative framework. Our analytical and numerical results show 
strong detrimental influence of the electron pairing on the effective 
antiferromagnetic coupling thereby suppressing the Kondo temperature 
in agreement with the experimental observations.

\end{abstract}  


\maketitle

{\em Motivation --} 
The recent electron tunneling experiments on the self-assembled quantum 
dots \cite{Deacon-10}, semiconducting nanowires \cite{Lee-12,Aguado-13} 
and/or carbon nano\-tubes \cite{Pillet-13,Schindele-13} coupled to one
superconducting and another conducting electrode provided evidence for 
the subgap bound states. They originate solely from the electron pairing 
which is spread onto nanoscopic objects activating the anomalous (Andreev) 
transport channel efficient even when the bias voltage $V$ is smaller 
than the energy gap $\Delta$ of superconductor. Similar in-gap states 
have been also detected \cite{Pillet-10,Dirks-11,Bretheau-13}  in the 
quantum dots connected to both superconducting reservoirs leading 
to inversion of the {\em dc} Josephson current ($0-\pi$ transition) 
\cite{Novotny-07}.

Correlated quantum dot (QD) coupled to the external conducting bath 
does usually induce the effective spin-exchange interactions, which 
(at low temperatures) may cause its total or partial screening. The 
resulting Kondo state shows up by a narrow Abrikosov-Suhl peak 
formed at the Fermi energy. For metallic junctions such effect  
has been predicted theoretically and observed experimentally
\cite{Pustilnik-04}, enhancing the zero-bias conductance.
In the metal - QD - superconductor (N-QD-S) heterostructures
the Kondo-type correlations are additionally confronted with   
electron pairing. Depending on the gate voltage, temperature
and  potential of the Coulomb repulsion  the QD ground state 
can vary from the (spinful) doublet $\left| \sigma \right>$
(where $\sigma=\uparrow,\downarrow$) to the (spinless)
 BCS configuration $u\left| 0 \right> - v \left| \uparrow
\downarrow\right>$ \cite{Bauer-07}. Such quantum phase 
transitions (showing qualitative influence of the induced 
on-dot pairing) can be experimentally observed in a tunable 
way \cite{Viewpoint-13}.  

Interplay between the superconductivity and Kondo-type 
correlations has been intensively explored experimentally  
\cite{Deacon-10,Lee-12,Aguado-13,Pillet-13,Schindele-13} and theoretically 
\cite{Rodero-11,Yamada-11_and_Bauer-13,theor-Andreev,Baranski-13}. 
From a physical point of view the most intriguing situation occurs, 
when the Kondo and proximity effects eventually coexist, leading  
to a tiny (yet clearly pronounced) enhancement  of the zero-bias 
subgap conductance reported independently by several groups 
\cite{Deacon-10,Aguado-13,Chang-13,Lohneysen-12}. Similar 
zero-bias feature is currently studied also for junctions 
made of $s$-wave superconductor coupled to quantum wires 
with the  strong spin-orbit interactions (for instance InSb or InAs) 
where the Majorana-type quasiparticles can appear \cite{Majorana-exp}.

In this paper we analyze the proximity induced pairing and study 
its influence on the antiferromagnetic exchange coupling (thereby 
on the Kondo temperature) using novel method based on the continuous 
unitary transformation (CUT). This technique is reminiscent of 
the renormalization group treatments and has a virtue to go beyond 
the perturbative scheme. Our study generalizes the famous Schrieffer-Wolf 
transformation \cite{S-W} by: i) considering the superconducting 
bath of itinerant electrons and ii) constructing the non-perturbative 
procedure reliable for the difficult case when the 
Kondo-type correlations compete with the induced on-dot pairing. 
This issue could be presntly of a broad interest for the nanoscopic, 
solid state and ultracold fermion atom communities. 

In the following we: 1) introduce the microscopic model of the proximized
quantum impurity, 2) construct the continuous canonical transformation 
expressing it through a set of the {\em flow} equations, 3) investigate 
the analytical (lowest order) solution, and 4) discuss the numerical 
results based on the selfconsistent Runge-Kutta algorithm. Our results 
reproduce the qualitative features obtained by the subgap tunneling 
spectroscopy \cite{Deacon-10,Lee-12,Aguado-13,Pillet-13,Schindele-13}.

{\em Proximity induced pairing --}
For studying a combined effect of the electron pairing and the 
Coulomb repulsion (which can induce the Kondo effect) we consider 
the Anderson impurity Hamiltonian
$\hat{H} = \sum_{\beta}\hat{H}_{\beta} + \sum_{\sigma}  
\varepsilon_{d} \hat{d}^{\dagger}_{\sigma} \hat{d}_{\sigma}  
+  U_{d}  \hat{n}_{ d \uparrow} \hat{n}_{ d \downarrow}  
+ \sum_{{\bf k},\sigma,\beta}  \left( V_{{\bf k} \beta} \; 
\hat{d}_{\sigma}^{\dagger}  \hat{c}_{{\bf k} \sigma \beta } 
+ V_{{\bf k} \beta}^{*}  \; \hat{c}_{{\bf k} \sigma, \beta }
^{\dagger} \hat{d}_{\sigma} \right)$. It formally describes 
the correlated quantum dot placed in between
the normal metal ($\beta\!=\!N$) and the superconducting 
($\beta\!=\!S$) electrodes. As usually, $\hat d_{\sigma}$ 
($\hat d_{\sigma}^{\dagger}$) denote the QD annihilation 
(creation) operators, $\sigma$ refers to spin $\uparrow$ or
$\downarrow$ configurations, $\varepsilon_{d}$ is the QD 
energy level, $U_{d}$ describes the repulsive Coulomb potential 
between the opposite spin electrons and  $V_{{\bf k}\beta}$
is the hybridization of the QD  electrons with external reservoirs. 

We treat electrons of the metallic reservoir as the free 
fermion gas $\hat{H}_{N} \!=\! \sum_{{\bf k},\sigma} 
\xi_{{\bf k}N}  \hat{c}_{{\bf k} \sigma N}^{\dagger} 
\hat{c}_{{\bf k} \sigma N}$  and describe the superconducting 
electrode by the BCS Hamiltonian $\hat{H}_{S} \!=\!\sum_{{\bf k},
\sigma}  \xi_{{\bf k}S}\hat{c}_{{\bf k} \sigma S }^{\dagger}  
\hat{c}_{{\bf k} \sigma S} \!-\! \sum_{\bf k} \Delta  \left( 
\hat{c}_{{\bf k} \uparrow S }^{\dagger} \hat{c}_{-{\bf k} 
\downarrow S }^{\dagger} + \hat{c}_{-{\bf k} \downarrow S} 
\hat{c}_{{\bf k} \uparrow S }\right)$. Energies of mobile
electrons $\xi_{{\bf k}\beta}\!=\!\varepsilon_{{\bf k}\beta} 
\!-\!\mu_{\beta}$ are measured with respect to the chemical 
potentials $\mu_{\beta}$ (which can be detuned by  voltage 
$V$ applied across the junction). In this work we focus on 
the equilibrium condition $\mu_{N}\!=\!\mu_{S}$ and the central 
task of our study is the effective low energy physics in a subgap 
regime $|\omega|<\Delta$. We shall assume the wide band 
limit approximation $|V_{{\bf k}\beta}| \!  \ll \! D$ (where $-D\!
\leq\!\varepsilon_{{\bf k}\beta} \!  \leq \! D$) and  
use the half-bandwidth $D$ as a convenient energy unit.
For simplicity we also impose the constant hybridization 
couplings $\Gamma_{\beta}\equiv 2\pi\sum_{{\bf k}}
|V_{{\bf k}\beta}|^{2}\delta(\omega-\xi_{{\bf k}\beta})$.   

Deep in the subgap regime $|\omega| \ll \Delta$ the electronic 
states are affected by the superconducting reservoir merely 
through the induced on-dot pairing gap $\Delta_{d}$. It can 
be shown (more detailed arguments are provided in section I of 
the supplementary material) that the strong hybridization 
$\Gamma_{S}$ induces the pairing gap $\Delta_{d} \simeq 
\Gamma_{S}/2$ \cite{Baranski-13}. Microscopic model 
of the proximized quantum dot can be thus represented by 
the auxiliary Hamiltonian
\begin{eqnarray}
\hat{H} &=& \sum_{{\bf k} \sigma} \xi_{\bf k} 
\hat{c}_{{\bf k}\sigma}^{\dagger} \hat{c}_{{\bf k}\sigma}
+ \sum_{\sigma} \varepsilon_{d} \hat{d}_{\sigma}^{\dagger} 
\hat{d}_{\sigma} - \Delta_{d}  \left( \hat{d}_{\uparrow}
^{\dagger} \hat{d}_{\downarrow}^{\dagger} + 
\hat{d}_{\downarrow}\hat{d}_{\uparrow} \right) 
\nonumber \\ &+& U_{d}{\hat n}_{ d \uparrow} {\hat n}_{d \downarrow} 
+ \frac{1}{\sqrt{N}}\sum_{{\bf k} \sigma} V_{\bf k} \left( 
\hat{c}_{{\bf k}\sigma}^{\dagger} \hat{d}_{\sigma} +
\hat{d}_{\sigma}^{\dagger} \hat{c}_{{\bf k} \sigma} 
\right) .  \label{hamil}
\end{eqnarray}
From now onwards we consider this Hamiltonian (\ref{hamil}) 
trying to determine the effective low energy physics in  
presence of correlations. 
Since we have to deal only  with the metallic reservoir we can 
abbreviate the notation by skipping the subindex $N$ in 
$\xi_{{\bf k}N}$, $V_{{\bf k}N}$ and 
$\hat{c}_{{\bf k}\sigma,N}^{(\dagger)}$.

{\em Outline of the CUT method --}
We shall now construct the  unitary transformation simplifying 
the model Hamiltonian (\ref{hamil}) to its equivalent easier form. 
Instead of single step transformation we use the novel method 
introduced  by F.\ Wegner \cite{Wegner-94} and independently by K.G.\ 
Wilson with S.\ G\l azek \cite{Wilson-94}. The underlying idea is 
a continuous transformation $\hat{H}(l) = \hat{\cal {U}}(l) 
\hat{H} \hat{\cal {U}}^{-1}(l)$ which via sequence of infinitesimal 
steps $l \rightarrow l+\delta l$ transforms the Hamiltonian to 
the required (diagonal, block-diagonal or any other) structure. 
Such continuous transformation depends on a specific 
choice of the operator $\hat{\cal {U}}(l)$. The transformed 
Hamiltonian obeys the {\em flow} equation
$\frac{ d  \hat {H}(l)}{dl}=\frac{d\hat{\cal {U}}(l)}{dl} 
\hat{H} \hat{\cal {U}}^{-1}(l)+\hat{\cal {U}}(l) 
\hat{H} \frac{d\hat{\cal {U}}^{-1}(l)}{dl}$, and due to the
identity $\hat{\cal {U}}(l) \hat{\cal {U}}^{-1}(l)=1$
implying $\frac{d \hat{\cal {U}}(l)}{dl} \hat{\cal {U}}^{-1}(l)
=-\hat{\cal {U}}(l) \frac{d\hat{\cal {U}}^{-1}(l)}{dl}$,
it can be formally expressed as follows \cite{Wegner-94}
\begin{eqnarray}
\frac{ d  \hat {H}(l)}{dl} = [\hat{\eta} (l), \hat {H}(l)] 
\label{General_flow}
\end{eqnarray}
with the generating operator $\hat {\eta} (l) \equiv \frac 
{d \hat{{\cal {U}}} (l)} {dl} \hat {{\cal {U}}}^{-1} (l)$.

The differential flow equation (\ref{General_flow}) enforces scaling 
(renormalization) of the model parameters (all quantities become
$l$-dependent). Initially mainly the large energy states are 
transformed whereas the small energy sector is rescaled 
later on \cite{Kehrein_book}. This continuous scaling proceeds, 
however, in the full Hilbert space. We thus keep information about 
all energy states and can study mutual feedback effects between 
the large and small energy sectors instead of integrating out 
'the fast modes'  typical for the RG methods.

The continuous transformation  of $\hat{H}(l)$ is controlled 
via equation (\ref{General_flow}) by the operator $\hat{\eta}(l)$. 
It has been shown by Wegner \cite{Wegner-94} that for Hamiltonian
$\hat{H}(l) = \hat{H}_{0}(l) + \hat{V}(l)$ it is convenient to 
choose
\begin{eqnarray}
\hat {\eta} (l) = \left [\hat {H}_{0}(l), \hat V (l) \right ]
\label{General_eta}
\end{eqnarray}
because (\ref{General_eta}) guarantees that ${\hat V}(l)$ vanishes 
in the asymptotic limit $ l \rightarrow \infty$ . Of course, there 
are possible also alternative options \cite{Kehrein_book}. Our present 
study is based on the scheme (\ref{General_eta}). We would like to remark 
that CUT method has been already successfully applied to the single 
impurity Anderson model (in absence of the proximity induced 
pairing) by S.\ Kehrein and A.\ Mielke \cite{Kehrein-94}, 
revisiting the single step Schrieffer-Wolff (S-W) transformation 
\cite{S-W}. The authors have shown that the momentum dependent 
spin-exchange coupling is free of any divergences and close
to Fermi surface becomes  antiferromagnetic.

{\em The flow equations --}
We shall formulate a continuous extension of the S-W transformation 
for the Hamiltonian (\ref{hamil}) using  the choice 
(\ref{General_eta}) in order to eliminate the hybridization term
$\hat{V}(l) =\frac{1}{\sqrt{N}}\sum_{{\bf k} \sigma}  V_{\bf k}(l) 
\left( \hat{c}_{{\bf k}\sigma}^{\dagger} \hat{d}_{\sigma} +
\hat{d}_{\sigma}^{\dagger} \hat{c}_{{\bf k} \sigma} \right)$.  
During this process the paramters of $\hat{H}_{0}(l) \equiv 
\hat{H}(l)-\hat{V}(l)$ are renormalized and some additional 
terms become generated (see section II of the Supplementary material). The generating 
operator (\ref{General_eta}) has  antihermitean  structure $\hat{\eta}(l)
=\hat{\eta}_{0}(l)-\hat{\eta}_{0}^{\dagger}(l)$, where
\begin{eqnarray}
&&\hat{\eta}_{0}(l) = \sum_{{\bf k} \sigma} \left ( \eta_{\bf k}(l) +
\eta_{\bf k}^{(2)}(l) \hat{d}^{\dagger}_{-\sigma} \hat{d}_{-\sigma}
\right)\hat{c}_{{\bf k}\sigma}^{\dagger} \hat{d}_{\sigma} 
\label{eta} \\ && 
+ \sum_{{\bf k p} \sigma } \eta_{\bf k p}(l) \hat{c}_{{\bf k}
\sigma}^{\dagger} \hat{c}_{{\bf p}\sigma} + \sum_{\bf k} 
\eta^{(1)}_{\bf k}(l) \left( \hat{c}_{{\bf k} \uparrow}^{\dagger}
\hat{d}_{\downarrow}^{\dagger}  - \hat{c}_{{\bf k} \downarrow}
^{\dagger} \hat{d}_{\uparrow}^{\dagger}  \right) 
\nonumber
\end{eqnarray}
and $l$-dependent  coefficients are given by
$\eta_{\bf k}(l)= \frac{1}{\sqrt N}(\xi_{\bf k}(l)-\varepsilon_{d}(l)) 
V_{\bf  k}(l)$, $\eta_{\bf k p}(l)=\frac{1}{N}V_{\bf k}(l)V_{\bf p}$(l), 
$\eta_{\bf  k}^{(1)}(l)=\frac{1}{\sqrt N} \Delta_{d}(l) V_{\bf k}(l)$, 
$\eta_{\bf  k}^{(2)}(l)=-\frac{1}{\sqrt N}U(l) V_{\bf k}(l)$.
Let us remark  that the standard S-W transformation 
$e^{\hat{S}}\hat{H}e^{-\hat{S}}$ \cite{S-W} can be reproduced with
the operator $\hat{S}$ of the same structure as (\ref{eta}) using
$\eta_{\bf k}=\frac{1}{\sqrt{N}}V_{\bf k}/\left( \xi_{\bf k}
-\varepsilon_{d}\right)$, $\eta^{(2)}_{\bf k}=\frac{1}{\sqrt{N}}
V_{\bf k}U/\left[ \left( \varepsilon_{d} -\xi_{\bf k} \right) 
\left( \varepsilon_{d}+U-\xi_{\bf k}\right)\right]$ and  
$\eta_{\bf  k}^{(1)}=0=\eta_{\bf k p}$. This fact indicates 
common roots of the single step and continuous 
transformation for a given problem at hand.

Substituting (\ref{eta}) to the right h.s\ of the flow equation 
(\ref{General_flow}) we obtain some terms, which initially were
absent in the model Hamiltonian (\ref{hamil}). From these new 
contributions we take here into account only the spin-exchange 
interactions, essential for the Kondo physics (but this procedure 
can be easily extended on other interactions). We update the initial 
Hamiltonian (\ref{hamil}) by $H_{exch}(l) = -  \sum_{{\bf k}, 
{\bf p}} J_{\bf k p}(l) \hat{\bf s}_{d} \cdot \hat{\bf S}
_{{\bf k}{\bf p}}$ with the boundary constraint $J_{\bf k p}(0) =0$. 
Spin operator of the QD is denoted by $\hat{\bf s}_{d}$ and
$\hat{\bf S}_{{\bf k}{\bf p}}$ describes spins of mobile electrons 
of the metallic lead. From the lengthy but straightforward algebra 
(see the section II.b of the supplementary material) we derive 
the following set of coupled {\em flow equations} 
 \begin{eqnarray}
\frac {d \varepsilon_{d}(l)}{dl} &=& - \frac{2}{\sqrt N} \sum_{\bf k}  
\eta_{\bf k}(l) V_{\bf k}(l), \label{flow_epsd} \\
\frac {d U_{d}(l)}{dl} &=& - \frac{4}{\sqrt N} \sum_{\bf k} \eta_{\bf k}^{(2)}(l)
 V_{\bf k}(l) , \label{flow_U} \\
\frac {d \Delta_{d}(l)}{dl} &=& \frac{2}{\sqrt N} \sum_{\bf k} 
\eta_{\bf k}^{(1)}(l) V_{\bf k}(l) , \label{flow_Delta} \\
\frac {d V_{\bf k}(l)}{dl} &=& \eta_{\bf k}(l) \left  
[ \varepsilon_{d}(l)-\xi_{\bf k}(l) + U_{d}(l) \langle \hat{n}_{d,\sigma}
\rangle \right ] \label{flow_V} \\
&+& \frac{2}{\sqrt N} \sum_{\bf p} \eta_{\bf k p}(l) V_{\bf
  p}(l) - \eta_{\bf k}^{(1)}(l) \Delta_{d}(l)  \nonumber \\
&+& \eta_{\bf k}^{(2)}(l)
\left [\varepsilon_{d}(l)-\xi_{\bf k}(l) + U_{d}(l) \right ] 
\langle \hat{n}_{d,\sigma} \rangle , \nonumber \\
\frac {d J_{\bf k p}(l)}{dl} &=& \eta_{\bf k}^{(2)}(l)V_{\bf p}(l)
+ \eta_{\bf p}^{(2)}(l) V_{\bf k}(l) \nonumber \\
&-& ( \xi_{\bf k}-\xi_{\bf p})^2 J_{\bf k p}(l) . 
\label{flow_J}
\end{eqnarray}
We skipped the derivative $\frac{d}{dl}\xi_{\bf k}(l)$ because it 
vanishes in the thermodynamic limit $N \rightarrow \infty$, 
implying  $\xi_{\bf k}(l)=\xi_{\bf k}$.

{\em Lowest order estimation --}
To gain some analytical (although approximate) solution of
the flow equations (\ref{flow_epsd}-\ref{flow_J}) we use 
the lowest order iterative estimation, justified  for 
$V_{\bf k}<<D$. In the first step we estimate $V_{\bf k}(l)$ 
solving the equation (\ref{flow_V}) upon neglecting 
$l$-dependence of all other parameters. To simplify 
such analysis we restrict to  the half-filled quantum dot 
case $n_{d\sigma}=0.5$ (i.e.\ $\varepsilon_{d}=-U_{d}/2$). 
Neglecting the cubic term $\eta_{\bf k p}(l) V_{\bf p}(l)$ 
in (\ref{flow_V}) we obtain 
%
\begin{eqnarray}
V_{\bf k}(l)=V_{\bf k} \; \mbox{\rm exp}\left[ -f_{\bf k} l\right] ,
\label{V_estimated}
\end{eqnarray}
where $f_{\bf k}\equiv \left( \varepsilon_{d}-\xi_{\bf k}\right)^{2}+ 
\Delta_{d}^{2} + \left( \varepsilon_{d} +U_{d}/2 - \xi_{\bf k}\right)U_{d}$. 
This expression (\ref{V_estimated}) yields an exponential disappearance 
of the hybridization coupling $V_{\bf k}(l)$. In the next step way 
can estimate $l$-dependence of all other quantities. Since we are 
particularly interested in the spin interactions we provide  
explicit expression only for the exchange coupling  
\begin{eqnarray}
J_{\bf k p}(l) &=& \frac{-2U_{d}V_{\bf k}V_{\bf p}}{f_{\bf k}\!+\!f_{\bf p}
\!-\!\left( \xi_{\bf k}\!-\!\xi_{\bf p}\right)^{2}}
\left[ 1 - e^{-\left(f_{\bf k}+f_{\bf p}\right) l}\right] .
\label{J_estimated} 
\end{eqnarray}
Nearby the Fermi surface (when $\xi_{{\bf k}_{F}}\!=\!0\!=\!
\xi_{{\bf p}_{F}}$) the exchange coupling (\ref{J_estimated}) 
becomes negative (antiferromagnetic), approaching the following 
asymptotic value
\begin{eqnarray}
J_{{\bf k}_F {\bf p}_F}(l \rightarrow \infty)
= \frac {- 4 U_{d} |V_{{\bf k}_F}|^2}{U_{d}^{2} + 
\left(2\Delta_{\bf d}\right)^2}.
\label{lowest}
\end{eqnarray}
Let us compare this result (\ref{lowest}) to the value 
$- 4 |V_{{\bf k}_F}|^2/U_{d}$ obtained previously from the 
S-W \cite{S-W} and the CUT study \cite{Kehrein-94} for 
the half-filled Anderson impurity embedded in a metallic medium. 
We  notice that the proximity induced on-dot pairing 
$\Delta_{d}$ substantially weakens the exchange coupling. 

{\em Numerical solution --}
To check the validity of our analytical results we solved 
the flow equations (\ref{flow_epsd}-\ref{flow_J}) fully selfconsistently, 
implementing the numerical Runge-Kutta algorithm. For the computations 
we discretized the energy band $\xi_{\bf k}/D=-1+2|k|$ by a mesh 
of 1000 equidistant points $k\in[-1,1]$. For the half-filled QD the
 Fermi level $\xi_{{\bf k}_F}=0$ corresponds to $|k|=0.5$. We carried 
out the calculations for small hybridization $V_{\bf k}=D/10$ 
focusing on the half-filled quantum dot case $n_{d\sigma}=0.5$.
All $l$-dependent quantities were determined from the following
iterative scheme $x(l+\delta l)\simeq x(l)+x'(l)\delta l$ with
derivative $x'(l)$ taken from the flow equations 
(\ref{flow_epsd}-\ref{flow_J}). We changed the increment 
$\delta l$, depending on a magnitude the continuous parameter $l$. 
At initial steps of the transformation we used $\delta l =0.01$  
(for $0\leq l< 1)$ and gradually increased it for higher values of $l$
(these values are expressed in units $D^{-2}$). We continued the
numerical procedure calculating all $l$-dependent quantities up 
to $l=100$, when   $V_{\bf k}(l)$ decreased
more than $6$ orders from its initial value.  

\begin{figure}
\epsfxsize=9cm{\epsffile{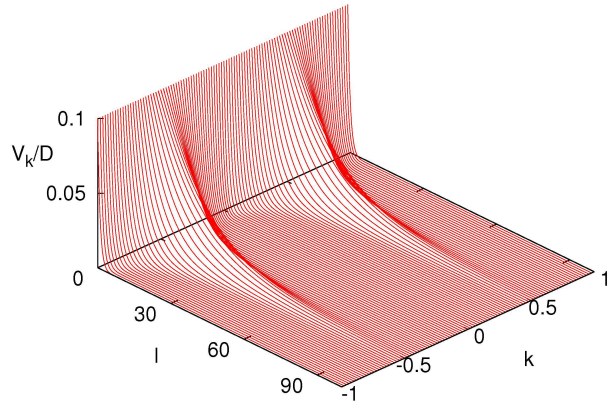}}
\caption{(color online) Flow of the hybridization coupling $V_{\bf k}(l)$ 
obtained for the initial values of the model (\ref{hamil}) parameters: 
$\varepsilon_{d}/D=-0.2$, $U_{d}/D=0.4$, $\Delta_{d}/D=0.1$ and 
$V_{\bf k}/D=0.1$.}
\label{Fig_Vk}
\end{figure}

Figure \ref{Fig_Vk} shows variation of the hybridization coupling 
$V_{\bf k}(l)$ with respect to the flow parameter $l$. We clearly see
that it vanishes, roughly obeying the exponential relation (\ref{V_estimated}). Hybridization of the electronic states distant from the Fermi level are
transformed pretty fast, whereas the states closer nearby the Fermi 
momentum $k_{F}=\pm 0.5$ are eliminated later on. This procedure resembles 
integrating out the fast and slow energy modes by the numerical 
renormalization group methods.

\begin{figure}
\epsfxsize=6.5cm{\epsffile{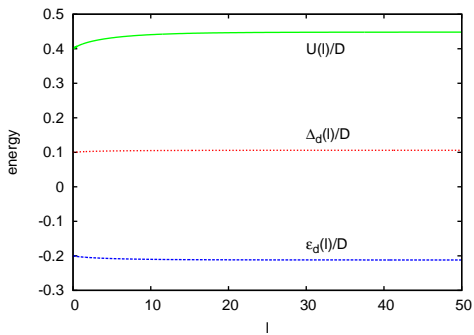}}
\caption{(color online) Variation of the quantum dot energy $\varepsilon_{d}(l)$, 
Coulomb repulsion $U_{d}(l)$, and the on-dot pairing $\Delta_{d}(l)$ 
with respect to $l$ for the same set of parameters as in fig.\ 
\ref{Fig_Vk}.}
\label{Fig_scaling}
\end{figure}

Exponential decrease of $V_{\bf k}(l)$ is accompanied by ongoing 
renormalization of the QD energy $\varepsilon_{d}(l)$, Coulomb
interaction $U_{d}(l)$ and the pairing gap $\Delta_{d}(l)$. Since 
we assumed the hybridization to be small therefore these renormalizations 
proved to be rather marginal (figure \ref{Fig_scaling}). 

\begin{figure}
\epsfxsize=9cm{\epsffile{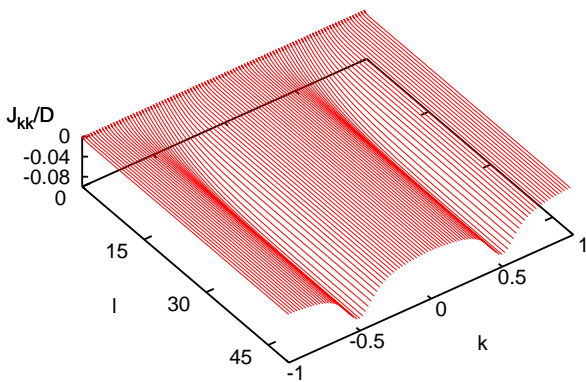}}
\caption{(color online) The spin exchange coupling $J_{\bf k p}
(l)$ obtained for the same set of parameters as in figure 
\ref{Fig_scaling}. For illustration we choose 
${\bf k}\!=\!{\bf p}$, when the exchange coupling is negative 
(otherwise $J_{{\bf k}\!\neq\!{\bf p}}(l)$ changes the sign 
for momenta distant from the Fermi surface).}
\label{Fig_Jkk}
\end{figure}

\begin{figure}
\epsfxsize=8cm{\epsffile{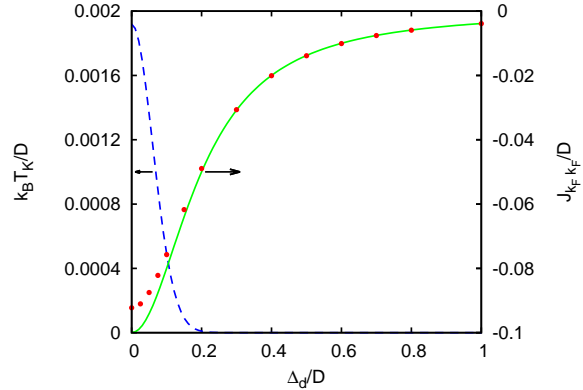}}
\caption{The asymptotic limit $l\rightarrow \infty$ value of the effective 
exchange coupling $J_{{\bf k}_F {\bf k}_F}$ as a function of the on-dot energy 
gap $\Delta_{d}$ obtained for the symmetric Anderson impurity with the following 
(initial) parameters:  $V_{\bf k}/D=0.1$, $\varepsilon_{\bf d}/D=-0.2$ and 
$U=-2\varepsilon_{d}$. The results based on the lowest order estimation (\ref{lowest}) 
(solid line) nearly coincide with the fully selfconsistent numerical solution 
(points). The dashed curve shows the corresponding Kondo temperature $k_B T_K$.}
\label{Vkkdd}
\end{figure}

The most important physical result is the induced spin-exchange 
coupling $J_{{\bf k}{\bf p}}(l)$. Figure \ref{Fig_Jkk} illustrates 
its $l$-dependence obtained for ${\bf k}\!=\!{\bf p}$. We can 
notice the negative (antiferromagnetic) coupling which is strongly 
enhanced nearby the Fermi surface, in agreement with (\ref{J_estimated}). 
We repeated the selfconsistent numerical calculations for a number
of $\Delta_{d}$ values. The effective (asymptotic  limit) value 
$J_{{\bf k}_F {\bf p}_F} (l\!=\!\infty)$ is shown by points in 
figure \ref{Vkkdd}. For comparison we also plot the analytical 
value (solid line). The analytical formula 
(\ref{lowest}) overestimates  $J_{{\bf k}_F {\bf p}_F} (l\!=\!
\infty)$ by a few percent. Summarizing, we 
conclude that the induced on-dot pairing has a detrimental 
influence on the antiferromagnetic coupling. To get some insight 
into the Kondo temperature $T_{K}$ we estimate its value from  the
Bethe-ansatz formula  \cite{Tsvelik-83}
$k_{B} T_{K} = \frac{2}{\pi} D \; \mbox{\rm exp}
\left\{ -\phi \left [ 2\rho(\varepsilon_{\bf F}) 
J_{{\bf k}_F {\bf p}_F} (l\!=\!\infty) \right ]\right\}$,
where $\rho(\varepsilon_{F}$) is the density of states at 
the Fermi level and $\phi(y) \simeq |y|^{-1} - 0.5 \ln{|y|}$. 
The obtained Kondo temperature is plotted by dashed line in figure 
\ref{Vkkdd}. We notice that $T_{K}$ is strongly suppressed 
by the on-dot pairing $\Delta_{d}$, reproducing qualitatively 
the experimental results \cite{Lee-12}.

{\em Summary and outlook --}
We have investigated the correlated quantum dot coupled between 
the superconducting and metallic reservoirs. Using the continuous 
unitary transformation we have determined the effective spin-exchange 
coupling  between the QD and metallic electrons. At the Fermi 
level such interactions have antiferromagnetic character which is
necessary for inducing the Kondo effect. Our approximate analytical 
formula (\ref{lowest}) and the fully selfconsistent numerical 
solution of the flow equations show that the on-dot pairing 
$\Delta_{d}$ substantially weakens such antiferromagnetic 
exchange coupling. In consequence, the Kondo temperature is 
strongly suppressed by the induced on-dot pairing. This behavior
has been indeed observed experimentally (where effective pairing gap 
was modified by the magnetic field) \cite{Lee-12}.

Further extension of the present study could be worthwhile for 
the nonequilibrium situation $\mu_{N}\neq\mu_{S}$. To calculate 
the charge current (of the Andreev and other channels) one can adopt 
the scheme formulated for the  QD coupled between both metallic 
leads \cite{Fritsch-10_and_Wang-10}. Besides considering the 
many-body phenomena under nonequilibrium conditions 
\cite{other-flow} it could be also interesting to extend the present
study to multiterminal configurations with the superconducting
electrodes, where electrons released from the Cooper pairs 
preserve entanglement.  

We acknowledge  discussions with  R.\ Aguado, J.\ Bauer,  
V.\ Jani\v{s}, S.\ Kehrein, T.\ Novotn\'y, and K.I.\ Wysoki\'nski.

\end{document}


      \title{Supplementary material to 'Kondo impurity between 
              superconducting\\ and metallic reservoir: the flow 
              equation approach'}

     \author{M.\ Zapalska and T.\ Doma\'nski}
\affiliation{
             Institute of Physics, 
	     M.\ Curie Sk\l odowska University, 
             20-031 Lublin, Poland} 
      \date{\today}

\maketitle
\section{Hamiltonian of proximized QD}

For microscopic description of the N-QD-S heterojunction we 
can use the Anderson-type Hamiltonian 
\begin{eqnarray} 
\hat{H} &=& \sum_{\beta}\hat{H}_{\beta} + \sum_{\sigma}  
\epsilon_{d} \hat{d}^{\dagger}_{\sigma} \hat{d}_{\sigma}  
+  U_{d} \; \hat{n}_{d \uparrow} \hat{n}_{d \downarrow}  
\label{model} \\
&+& \sum_{{\bf k},\sigma}\sum_{{\beta}}  \left( V_{{\bf k} \beta} \; 
\hat{d}_{\sigma}^{\dagger}  \hat{c}_{{\bf k} \sigma \beta } 
+  V_{{\bf k} \beta}^{*}  \hat{c}_{{\bf k} \sigma 
\beta }^{\dagger} \hat{d}_{\sigma} \right) ,
\nonumber 
\end{eqnarray} 
where the subindex $\beta$ refers either to the normal metal ($N$)
or the superconducting ($S$) electrode. Reservoirs of such itinerant
electrons can be represented correspondingly by the Hamiltonian 
of a free Fermi gas 
$\hat{H}_{N} \!=\! \sum_{{\bf k},\sigma} \xi_{{\bf k}N}  
\hat{c}_{{\bf k} \sigma N}^{\dagger} \hat{c}_{{\bf k} \sigma N}$  
and the usual BCS form
$\hat{H}_{S} \!=\!\sum_{{\bf k},\sigma}  \xi_{{\bf k}S}
\hat{c}_{{\bf k} \sigma S }^{\dagger}  \hat{c}_{{\bf k} \sigma S} 
\!-\! \sum_{\bf k} \Delta  \left( \hat{c}_{{\bf k} \uparrow S }
^{\dagger} \hat{c}_{-{\bf k} \downarrow S }^{\dagger} + \hat{c}
_{-{\bf k} \downarrow S} \hat{c}_{{\bf k} \uparrow S }\right)$.

The qualitative features originating from the proximity effect 
can be deduced by studying the single particle Green's function 
${\mb G}_{d}(\tau,\tau')\!=\!\langle\langle \hat{\Psi}_{d}(\tau); 
\hat{\Psi}_{d}^{\dagger}(\tau')\rangle\rangle$ in the Nambu 
spinor representation $\hat{\Psi}_{d}^{\dagger}=(\hat{d}_
{\uparrow}^{\dagger},\hat{d}_{\downarrow})$, $\hat{\Psi}_{d}
=(\hat{\Psi}_{d}^{\dagger})^{\dagger}$. In absence of an external 
voltage the Green's function ${\mb G}_{d}(\tau,\tau')$ depends only 
on time difference $\tau\!-\!\tau'$ and its Fourier transform obeys
\begin{eqnarray} 
\left[{\mb G}_{d}(\omega)\right]^{-1} = 
\left( \begin{array}{cc}  
\omega\!-\!\varepsilon_{d} &  0 \\ 0 &  
\omega\!+\!\varepsilon_{d}\end{array}\right)
- {\mb \Sigma}_{d}^{0}(\omega)  
- {\mb \Sigma}_{d}^{U}(\omega) .  
\label{GF}\end{eqnarray} 
The first contribution  ${\mb \Sigma}_{d}^{0}$ takes into account the 
hybridization effects (of an uncorrelated quantum impurity) whereas
the second part ${\mb  \Sigma}_{d}^{U}$ describes the corrections 
due to the Coulomb repulsion $U_{d}$. 

\begin{figure}
\epsfxsize=8cm\centerline{\epsffile{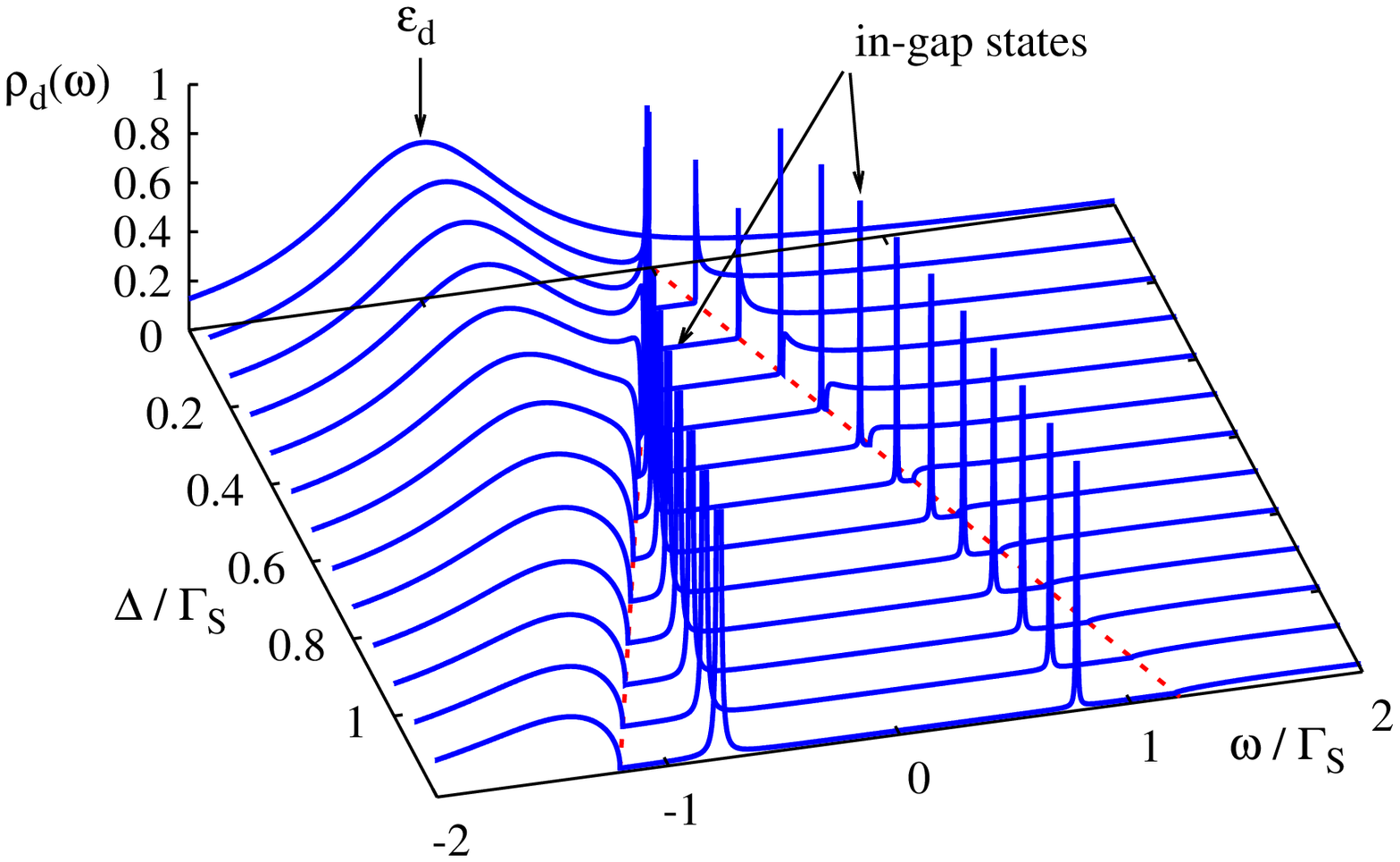}}
\vspace{0.0cm}
\caption{Spectral function $\rho_{d}(\omega)$ of 
the uncorrelated quantum dot obtained for $\varepsilon_{d}/\Gamma_{S}=-1$ 
assuming a strong asymmetry of the hybridization $\Gamma_{N}/\Gamma_{S}=10^{-3}$. 
We can notice that the in-gap (Andreev) quasiparticles emerge from the 
singularities $\pm \Delta$ (dashed lines) and  gradually evolve to $\pm 
\sqrt{\varepsilon_{d}^{2}+(\Gamma_{S}/2)^{2}}$. For arbitrary 
parameters they appear symmetrically around the Fermi level 
(chosen here as $\omega=0$).}
\label{proximity_effect}
\end{figure}

The hybridization part $\mb{\Sigma}_{d}^{0}(\omega)$ is known exactly. 
Its explicit form for the wide-band limit is found as \cite{Bauer-07} 
\begin{eqnarray}
\mb{\Sigma}_{d}^{0}(\omega) &=&  - \; \frac{\Gamma_{N}}{2} \; 
\left( \begin{array}{cc}  
i & 0 \\ 0 & i \end{array} \right)
\label{selfenergy_0} \\
& - & \frac{\Gamma_{S}}{2} \left( \begin{array}{cc}  
1 & \frac{\Delta}{\omega} \\ 
 \frac{\Delta}{\omega}  & 1 
\end{array} \right) \times
\left\{
\begin{array}{ll} 
\frac{\omega}{\sqrt{\Delta^{2}-\omega^{2}}}
& \mbox{\rm for }  |\omega| < \Delta , \\
\frac{i\;|\omega|}{\sqrt{\omega^{2}-\Delta^{2}}}
& \mbox{\rm for }  |\omega| > \Delta .
\end{array} \right. 
\nonumber
\end{eqnarray} 
Roughly speaking, the selfenergy $\mb{\Sigma}_{d}^{0}(\omega)$ is  
responsible for: a) the induced on-dot pairing (due to off-diagonal
terms which are proportional to $\Gamma_{S}$) and b) the finite 
life-time effects (i.e.\ broadening of the QD states). The latter 
effect depends either on both couplings $\Gamma_{\beta =N,S}$ 
(for energies $|\omega|\geq\Delta$) or solely on $\Gamma_{N}$ 
(in a subgap regime  $|\omega|<\Delta$). Figure 
\ref{proximity_effect} illustrates the typical spectral 
function $\rho_{d}(\omega)=-\pi^{-1}\mbox{\rm Im}{\mb G}_{d}
(\omega+i0^{+})$ obtained for the uncorrelated ($U_{d}\!=\!0$) 
quantum dot.

In a subgap regime $|\omega|<\Delta$ the single particle Green's 
function of the uncorrelated quantum dot has the BCS-type 
structure
\begin{eqnarray}
{\mb G}_{d}(\omega) &=& 
\left( \begin{array}{cc}  
\tilde{\omega} + \varepsilon_{d} + i \Gamma_{N}/2  
\hspace{0.3cm}&  \tilde{\Gamma}_{s}/2 \\ 
\tilde{\Gamma}_{s}/2  & 
\tilde{\omega} - \varepsilon_{d} + i \Gamma_{N}/2  
\end{array} \right)^{-1} 
\label{G_0}
\end{eqnarray}
with  $\tilde{\omega} = \omega + \frac{\Gamma_{S}}{2} \frac{\omega}
{\sqrt{\Delta^{2}-\omega^{2}}}$ and $\tilde{\Gamma}_{s} = \Gamma_{S} 
\frac{\Delta}{\sqrt{\Delta^{2}-\omega^{2}}}$. The subgap spectrum is 
thus characterized by two quasiparticle peaks, widely known in the 
literature as the Andreev \cite{Bauer-07} or Shiba-Rusinov  
\cite{Yu-Shiba-Rusinov} states. Let us emphasize that they originate 
from the induced on-dot pairing.

\begin{figure}
\epsfxsize=9cm\centerline{\epsffile{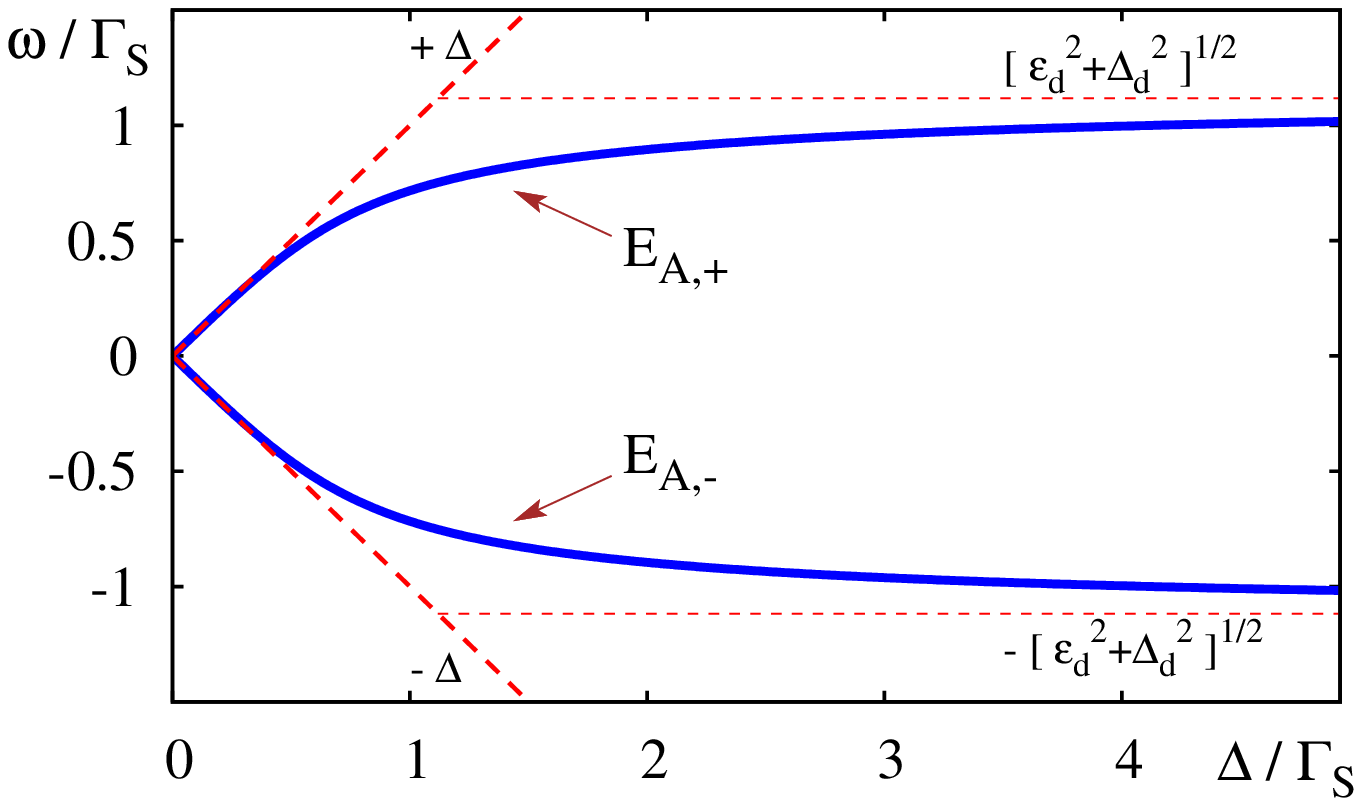}}
\vspace{0.0cm}
\caption{Energies $E_{A,\pm}$ of the subgap
(Andreev) quasiparticles  for the same set of parameters as 
in figure \ref{proximity_effect}.}
\label{in-gap-energies}
\end{figure}

For infinitesimally small coupling $\Gamma_{N}$ the in-gap spectrum 
consists of the Dirac-deltas (corresponding to the long-lived 
quasiparticles). Otherwise the Andreev states acquire some finite 
broadening, roughly controlled by the hybridization $\Gamma_{N}$. 
The quasiparticle energies $E_{A,\pm}$ of the uncorrelated quantum 
dot can be determined from the following relation \cite{Baranski-13}
\begin{eqnarray}
E_{A,\pm} +  \frac{(\Gamma_{S}/2)E_{A,\pm}} {\sqrt{\Delta^{2}
-E_{A,\pm}^{2}}} = \pm \sqrt{\varepsilon_{d}^{2}+ 
\frac{(\Gamma_{S}/2)^{2}\Delta^{2}}{\Delta^{2}-E_{A,\pm}^{2}}} .
\label{energy_eqn}
\end{eqnarray}
In figure \ref{in-gap-energies} we plot these quasiparticle energies 
$E_{A,\pm}$ versus the superconducting gap $\Delta$ for $\varepsilon
_{d}/\Gamma_{S}=-1$, $U_{d}\!=\!0$, $\Gamma_{N}\ll\Gamma_{S}$. 
In the limit $\Delta \ll \Gamma_{S}$ the Andreev states are located 
close-by the gap edge singularities $E_{A,\pm}\simeq \pm\Delta$. 
In the other extreme limit $\Delta \gg \Gamma_{S}$ they asymptotically 
approach $E_{A,\pm} \simeq \pm \sqrt{\varepsilon_{d}+\left(\Gamma_{S}/2
\right)^{2}}$. In the latter case (known as {\em the superconducting 
atomic limit}) the hybridization selfenergy (\ref{selfenergy_0}) 
simplifies to the static value
\begin{eqnarray}
\mb{\Sigma}_{d}^{0}(\omega) &=&  - \; \frac{1}{2} \; 
\left( \begin{array}{cc}  i\Gamma_{N} & \Gamma_{S} \\ 
\Gamma_{S} & i\Gamma_{N} \end{array} \right) .
\end{eqnarray}
One can hence replace the initial Hamiltonian (\ref{model})
by its equivalent  version 
\begin{eqnarray} 
\hat{H} &=&  \sum_{\sigma}  
\epsilon_{d} \hat{d}^{\dagger}_{\sigma} \hat{d}_{\sigma}  
+  U_{d} \; \hat{n}_{d \uparrow} \hat{n}_{d \downarrow}  
- \left( \Delta_{d} \hat{d}_{\uparrow}^{\dagger} 
\hat{d}_{\downarrow}^{\dagger} + \mbox{\rm h.c.} \right) 
\nonumber \\
&+& \hat{H}_{N} + \sum_{{\bf k},\sigma}  \left( V_{{\bf k} N} \; 
\hat{d}_{\sigma}^{\dagger}  \hat{c}_{{\bf k} \sigma N} 
+  V_{{\bf k} N}^{*}  \hat{c}_{{\bf k} \sigma 
N}^{\dagger} \hat{d}_{\sigma} \right) , 
\label{proximized} 
\end{eqnarray} 
where a role of the superconducting electrode is played  
by the induced on-dot gap $\Delta_{d}=\Gamma_{S}/2$.

\section{Continuous S-W transformation}

In this section we present some details on the continuous Schrieffer-Wolf 
transformation $\hat{H}(l) = \hat{\cal {U}}(l) \hat{H} \hat{\cal 
{U}}^{\dagger}(l)$ for the Hamiltonian (\ref{proximized}). Here and 
in the main text of the paper  we abbreviate the notation skipping 
the subindex $N$ in $V_{{\bf k} N}$ and  $\xi_{{\bf k} N}$. We are
going to construct such unitary transformation $\hat{\cal {U}}(l)$ 
eliminating the hybridization term 
$\hat{V}(l)=\sum_{{\bf k},\sigma}  \left( V_{{\bf k}}(l) \; 
\hat{d}_{\sigma}^{\dagger}  \hat{c}_{{\bf k} \sigma} 
+  V^{*}_{{\bf k}}(l)  \hat{c}_{{\bf k} \sigma}^{\dagger} 
\hat{d}_{\sigma} \right)$ in the asymptotic limit 
$\lim_{l\rightarrow\infty}V_{\bf k}(l)=0$.

\subsection{The generating operator}

To derive the effective Hamiltonian $\hat{H}(l\rightarrow\infty)$
we follow strictly the method introduced by Wegner \cite{Wegner-94}.
Evolution of the Hamiltonian has to be derived from the formal
differential equation
\begin{eqnarray}
\frac{ d  \hat {H}(l)}{dl} = [\hat{\eta} (l), \hat {H}(l)] ,
\label{hamil_flow}
\end{eqnarray}
where the generating operator is defined by $\hat{\eta}(l) 
\equiv \frac { d\hat{{\cal {U}}}(l)}{dl} \hat{{\cal {U}}}^{-1}(l)$.
According to Ref.\ \cite{Wegner-94} one possible way for choosing
$\hat{\eta}(l)$ is
\begin{eqnarray}
\hat{\eta}(l) = \left[ \hat{H}(l),\hat{V}(l) \right] 
\label{General_eta}
\end{eqnarray}
although also other alternative options are available 
\cite{Kehrein_book}.  
For the considered Hamiltonian (\ref{proximized}) the canonical 
operator $\hat{\eta}(l)$ has the  explicit form

\begin{eqnarray}
\hat \eta(l) &=& \sum_{{\bf k} \sigma}  \left( \eta_{\bf k}(l) 
\hat{c}_{{\bf k}\sigma}^{\dagger} \hat{d}_{\sigma} - \mbox{\rm h.c.}
\right) \nonumber \\ &+& \sum_{{\bf k p} \sigma } \left( \eta_{\bf k p}(l) 
\hat{c}_{{\bf k}\sigma}^{\dagger} \hat{c}_{{\bf p}\sigma} 
- \mbox{\rm h.c.} \right) \nonumber \\
&+& \sum_{\bf k}  \left( \eta^{(1)}_{\bf k}(l) \left( 
\hat{c}_{{\bf k} \uparrow}^{\dagger} \hat{d}_{\downarrow}^{\dagger} 
- \hat{c}_{{\bf k}  \downarrow}^{\dagger}\hat{d}_{\uparrow}^{\dagger} 
\right) - \mbox{\rm h.c.} \right) \nonumber \\
&+& \sum_{{\bf k} \sigma} \eta_{\bf k}^{(2)}(l) \left( 
\hat{c}_{{\bf k} \sigma}^{\dagger} \hat{d}_{-\sigma}^{\dagger}
\hat{d}_{-\sigma}\hat{d}_{\sigma} - \mbox{\rm h.c.} \right)
\label{eta_wegner}
\end{eqnarray}
with  the following $l$-dependent coefficients
$\eta_{\bf k}(l)= \frac{1}{\sqrt N}(\xi_{\bf k}(l)-\varepsilon_{d}(l)) 
V_{\bf  k}(l)$, $\eta_{\bf k p}(l)=\frac{1}{N}V_{\bf k}(l)V_{\bf p}(l)$, 
$\eta_{\bf  k}^{(1)}(l)=\frac{1}{\sqrt N} \Delta_{d}(l) V_{\bf k}(l)$ 
and $\eta_{\bf  k}^{(2)}(l)=-\frac{1}{\sqrt N}U(l) V_{\bf k}(l)$.
Previous studies \cite{Kehrein-96} of the usual Anderson impurity
Hamiltonian (i.e.\ $\Delta_{d}=0$) have been done using the same 
generating operator (\ref{eta_wegner}) without the coefficient 
$\eta_{\bf k}^{(1)}(l)$.

\subsection{Derivation of the flow equations}

Substituting (\ref{eta_wegner}) to the flow equation (\ref{hamil_flow}) 
for the model Hamiltonian (\ref{proximized}) one obtains 
%
\begin{widetext}
\begin{eqnarray}
\left [\hat{\eta}(l),\hat{H}(l) \right ] &=&\frac{1}{\sqrt N} 
\sum_{\bf k p \sigma} \eta_{\bf k}^{(2)}(l) V_{\bf p}(l) \left( 
{\hat c}_{\bf k \sigma}^{\dagger}{\hat d}_{-\sigma}^{\dagger}
{\hat d}_{-\sigma} {\hat c}_{{\bf p} \sigma} - {\hat c}_{{\bf k} 
\sigma}^{\dagger}{\hat  d}_{-\sigma}^{\dagger}{\hat d}_{\sigma} 
{\hat c}_{{\bf p} - \sigma} + {\hat c}_{{\bf p} \sigma}^{\dagger}
{\hat d}_{-\sigma}^{\dagger}{\hat d}_{-\sigma} {\hat c}_{{\bf k} \sigma}
- {\hat c}_{{\bf p}- \sigma}^{\dagger}{\hat d}_{\sigma}^{\dagger}
{\hat d}_{- \sigma} {\hat c}_{{\bf k} \sigma} \right)  \nonumber \\
&+& \sum_{\bf k p \sigma} \left[\frac{1}{\sqrt{N}}\eta_{\bf k}(l) 
V_{\bf p}(l) + \eta_{\bf k p}(\xi_{\bf p}(l) -\xi_{\bf k}(l) ) 
\right] \left( \hat{c}_{{\bf k}\sigma}^{\dagger}\hat{c}_{{\bf p}\sigma}
+ \hat{c}_{{\bf p}\sigma}^{\dagger}\hat{c}_{{\bf k}\sigma}\right ) 
- \frac{2}{\sqrt N} \sum_{\bf k \sigma} \eta_{\bf k}(l) V_{\bf k}(l)
\hat{d}_{\sigma}^{\dagger} \hat{d}_{\sigma} \nonumber \\
&-& \frac{2}{\sqrt N} \sum_{\bf k} \eta_{\bf k}^{(1)}(l) V_{\bf k}(l)
\left (\hat{d}_{\uparrow}^{\dagger} \hat{d}_{\downarrow}^{\dagger} +
\hat{d}_{\downarrow} \hat{d}_{\uparrow} \right) - \frac{2}{\sqrt N} 
\sum_{\bf k \sigma} 
\eta_{\bf k}^{(2)}(l) V_{\bf k}(l) \hat{d}_{\sigma}^{\dagger}
\hat{d}_{-\sigma}^{\dagger} \hat{d}_{-\sigma}\hat{d}_{\sigma} \nonumber \\
&+&\sum_{\bf k \sigma} \left \lbrace
\eta_{\bf k}(l) \left  [ \varepsilon_{d}(l)-\xi_{\bf k}(l) \right] 
+ \frac{2}{\sqrt N} \sum_{\bf p} \eta_{\bf k p}(l) V_{\bf
  p}(l) - \eta_{\bf k}^{(1)}(l) \Delta_{d}(l) \right \rbrace \left 
  ( \hat{c}_{{\bf k}\sigma}^{\dagger} \hat{d}_{\sigma} +
\hat{d}_{\sigma}^{\dagger} \hat{c}_{{\bf k} \sigma} \right ) \nonumber \\
&+& \sum_{\bf k} \left \lbrace \eta_{\bf k}^{(1)}(l) U_{d}(l) - \eta_{\bf
  k}^{(2)}(l)\Delta_{d}(l) \right \rbrace  \left [ \left ( 
\hat{d}_{\uparrow}^{\dagger} \hat{d}_{\uparrow} 
\hat{d}_{\downarrow}^{\dagger} \hat{c}_{\bf k \uparrow}^{\dagger} -
\hat{d}_{\uparrow}^{\dagger} \hat{c}_{\bf k \downarrow}^{\dagger} 
\hat{d}_{\downarrow}^{\dagger} \hat{d}_{\downarrow} \right )
 + \mbox{\rm h.c.} \right ]\nonumber \\
&+& \sum_{\bf k \sigma}\left \lbrace \eta_{\bf k}(l) U_{d}(l) + \eta_{\bf
  k}^{(2)}(l) \left [\varepsilon_{d} (l)- \xi_{\bf k}(l) + U_{d}(l) \right ] 
\right \rbrace \left ( \hat{c}_{{\bf k}\sigma}^{\dagger}
\hat{d}_{\sigma} \hat{d}_{-\sigma}^{\dagger} \hat{d}_{-\sigma} 
+ \mbox{\rm h.c.} \right )\nonumber \\
&-& \frac {1}{\sqrt N} \sum_{\bf k p}
\eta_{\bf k}^{(1)}(l) V_{\bf p}(l) \left [ \left 
( \hat{c}_{{\bf k}\uparrow}^{\dagger}
\hat{c}_{{\bf p}\downarrow}^{\dagger}+ \hat{c}_{{\bf p}\uparrow}^{\dagger}
\hat{c}_{{\bf k}\downarrow}^{\dagger} \right) + \mbox{\rm h.c.} \right ] 
- \frac{1}{\sqrt N} \sum_{\bf k p \sigma} \eta_{\bf k}^{(2)}(l) V_{\bf p}(l)
\left ( \hat{c}_{{\bf k}\sigma}^{\dagger}
\hat{c}_{{\bf p}-\sigma}^{\dagger} \hat{d}_{-\sigma} \hat{d}_{\sigma} 
+ \mbox{\rm h.c.} \right ) \nonumber \\
&-&\sum_{\bf k} \left \lbrace \eta_{\bf k}(l)\Delta_{d}(l) 
+ \eta_{\bf k}^{(1)}(l)\left [ \xi_{\bf k}(l) + \varepsilon_{d}(l) \right]
+\eta_{\bf k}^{(2)}(l)\Delta_{d}(l) \right \rbrace \left [
\left ( \hat{c}_{{\bf k} \uparrow}^{\dagger}\hat{d}_{\downarrow}^{\dagger}
+ \hat{d}_{\uparrow}^{\dagger}\hat{c}_{{\bf k}\downarrow}^{\dagger} \right) 
+ \mbox{\rm h.c.} \right ] .
\label{Ham_flow}
\end{eqnarray}
On the right hand side of (\ref{Ham_flow}) we can notice several terms, 
which were initially absent in the model Hamiltonian  (\ref{proximized}). 
Some of them could be eliminated by a suitable modification of the generating
operator $\hat{\eta}$ as it has been done for the usual Anderson model 
{\cite{Kehrein-96}}. In the present study, however, we assume that the 
hybridization $V_{\bf k}$ is much smaller than all other parameters. 
This assumption allows us to simplify (\ref{Ham_flow}) using the 
linearizations
\begin{eqnarray}
\left [\hat{\eta}(l),\hat{H}(l) \right ] &\approx &
\frac{1}{\sqrt N} \sum_{\bf k p} \eta_{\bf k}^{(2)}(l) 
V_{\bf p}(l) \left( {\hat c}_{\bf k \uparrow}^{\dagger}
{\hat d}_{\bf \downarrow}^{\dagger}{\hat d}_{\downarrow} 
{\hat c}_{\bf p \uparrow} - {\hat c}_{\bf k \downarrow}^{\dagger}
{\hat  d}_{\uparrow}^{\dagger}{\hat d}_{\bf \downarrow} 
{\hat c}_{\bf p \uparrow} + {\hat c}_{\bf p \uparrow}^{\dagger}
{\hat d}_{\bf \downarrow}^{\dagger}{\hat d}_{\bf \downarrow} 
{\hat c}_{\bf k \uparrow}  - {\hat c}_{\bf p \downarrow}^{\dagger}
{\hat d}_{\bf \uparrow}^{\dagger}{\hat d}_{\bf \downarrow}  
{\hat c}_{\bf k \uparrow} \right ) \nonumber \\
&-& \frac{2}{\sqrt N} \sum_{\bf k} \eta_{\bf k}^{(1)}(l) V_{\bf k}(l)
\left (\hat{d}_{\uparrow}^{\dagger} \hat{d}_{\downarrow}^{\dagger} +
\hat{d}_{\downarrow} \hat{d}_{\uparrow} \right ) - \frac{2}{\sqrt N} 
\sum_{\bf k \sigma} \eta_{\bf k}^{(2)}(l) V_{\bf k}(l) 
\hat{d}_{\sigma}^{\dagger}\hat{d}_{-\sigma}^{\dagger} 
\hat{d}_{-\sigma}\hat{d}_{\sigma} \nonumber \\
&+&\sum_{\bf k \sigma} \left \lbrace \eta_{\bf k}(l) 
\left[ \varepsilon_{d}(l)-\xi_{\bf k}(l)+U_{d}(l) 
\langle \hat{n}_{d,-\sigma} \rangle \right] 
+ \frac{2}{\sqrt N} \sum_{\bf p} \eta_{\bf k p}(l) V_{\bf  p}(l) 
- \eta_{\bf k}^{(1)}(l) \Delta_{d}(l) \right. \nonumber \\
&+& \left. \eta_{\bf k}^{(2)}(l)
\left [\varepsilon_{d}(l)-\xi_{\bf k}(l) + U_{d}(l) \right] 
\langle \hat{n}_{d,-\sigma} \rangle \right \rbrace 
\left( \hat{c}_{{\bf k}\sigma}^{\dagger} \hat{d}_{\sigma} 
+ \hat{d}_{\sigma}^{\dagger} \hat{c}_{{\bf k} \sigma} \right) 
+ \hat{O}(l)
\label{H_jj_flow}
\end{eqnarray}
with the expectation value $\langle \hat{n}_{d, \sigma} \rangle
= \langle \hat{d}^{\dagger}_{\sigma} \hat{d}_{\sigma} \rangle $ 
and the higher order term $\hat{O}(l)$ defined as
\begin{eqnarray}
\hat{O}(l)= &-& \sum_{\bf k} \left \lbrace \eta_{\bf k}(l)\Delta_{d}(l) 
+ \eta_{\bf k}^{(1)}(l)\left [ \xi_{\bf k}(l) + \varepsilon_{d}(l)
+ U_{d} \langle {\hat n}_{d,\uparrow} \rangle \right ]
+\eta_{\bf k}^{(2)}(l)\Delta_{d}(l)
(1-\langle {\hat n}_{d,\uparrow} \rangle) \right \rbrace
\left ( \hat{c}_{{\bf k} \uparrow}^{\dagger}\hat{d}_{\downarrow}^{\dagger}
+ \hat{d}_{\downarrow} \hat{c}_{{\bf k}\uparrow} \right) \nonumber \\ 
&-& \sum_{\bf k} \left \lbrace \eta_{\bf k}(l)\Delta_{d}(l) 
+ \eta_{\bf k}^{(1)}(l)\left [ \xi_{\bf k}(l) + \varepsilon_{d}(l)
+ U_{d} \langle {\hat n}_{d,\downarrow} \rangle \right ]
+\eta_{\bf k}^{(2)}(l)\Delta_{d}(l) 
(1-\langle {\hat n}_{d,\downarrow} \rangle) \right \rbrace
\left ( \hat{d}_{\uparrow}^{\dagger} \hat{c}_{\bf k\downarrow}^{\dagger}
+ \hat{c}_{\bf k \downarrow} \hat{d}_{\uparrow} \right) \nonumber \\
 &-& \frac {1}{\sqrt N} \sum_{\bf k p} \eta_{\bf k}^{(1)}(l) 
V_{\bf p}(l) \left [ \left( \hat{c}_{{\bf k}\uparrow}^{\dagger}
\hat{c}_{{\bf p}\downarrow}^{\dagger}+ \hat{c}_{{\bf p}\uparrow}^{\dagger}
\hat{c}_{{\bf k}\downarrow}^{\dagger} \right) + \mbox{\rm h.c.} \right ] 
- \frac{1}{\sqrt N} \sum_{\bf k p \sigma} \eta_{\bf k}^{(2)}(l) V_{\bf p}(l)
\left ( \hat{c}_{{\bf k}\sigma}^{\dagger}
\hat{c}_{{\bf p}-\sigma}^{\dagger} \hat{d}_{-\sigma} \hat{d}_{\sigma} 
+ \mbox{\rm h.c.} \right) .
\label{reszta}
\end{eqnarray}
\end{widetext}

We next update the initial Hamiltonian (\ref{proximized}) by
the spin-exchange interaction $-  \sum_{{\bf k}, {\bf p}} 
J_{\bf k p}(l) \hat{\bf s}_{d} \cdot \hat{\bf S}_{{\bf k}{\bf p}}$
imposing the initial condition $J_{\bf k p}(0) =0$. 
Neglecting the term (\ref{reszta}) we finally obtain 
the set of coupled differential equations
\begin{eqnarray}
\frac {d \varepsilon_{d}(l)}{dl} &=& - \frac{2}{\sqrt N} \sum_{\bf k}  
\eta_{\bf k}(l) V_{\bf k}(l) \nonumber \\
\frac {d U_{d}(l)}{dl} &=& - \frac{4}{\sqrt N} \sum_{\bf k} \eta_{\bf k}^{(2)}(l)
 V_{\bf k}(l) \nonumber \\
\frac {d \Delta_{d}(l)}{dl} &=& \frac{2}{\sqrt N} \sum_{\bf k} 
\eta_{\bf k}^{(1)}(l) V_{\bf k}(l) \nonumber \\
\frac {d V_{\bf k}(l)}{dl} &=& \eta_{\bf k}(l) \left  
[ \varepsilon_{d}(l)-\xi_{\bf k}(l) + U_{d}(l) \langle \hat{n}_{d,-\sigma}
\rangle \right ] \nonumber \\
&-&  \eta_{\bf k}^{(1)}(l) \Delta_{d}(l) + 
\frac{2}{\sqrt N} \sum_{\bf p} \eta_{\bf k p}(l) V_{\bf p}(l) \nonumber \\
&+& \eta_{\bf k}^{(2)}(l)
\left [\varepsilon_{d}(l)-\xi_{\bf k}(l) + U_{d}(l) \right ] \langle 
\hat{n}_{d,-\sigma} \rangle \nonumber \\
\frac {d J_{\bf k p}(l)}{dl} &=& \eta_{\bf k}^{(2)}(l)V_{\bf p}(l)
+ \eta_{\bf p}^{(2)}(l) V_{\bf k}(l) \nonumber \\
&-& ( \xi_{\bf k}-\xi_{\bf p})^2 J_{\bf k p}(l)  
\label{flow_eqn_jj}
\end{eqnarray}
The spin exchange coupling $J_{\bf kp}(l)$ appearing in the flow 
equation (\ref{flow_eqn_jj}) can be treated within the lowest 
order analytical estimation. In the asymptotic limit $l\rightarrow
\infty$ it  evolves to 
\begin{eqnarray}
J_{\bf k p}(\infty) &=& -2U_{d}V_{{\bf k}}V_{{\bf p}} \;
\left\{ \left( \varepsilon_{d}\!-\!\xi_{{\bf k}}\right)^{2}+
\left( \varepsilon_{d}\!-\!\xi_{{\bf p}}\right)^{2}
\right. \nonumber \\ &+& \left.
 \Delta_{d}^{2} \!-\! \left(  \xi_{{\bf k}}\!+\!\xi_{{\bf p}}
\right)U_{d}\!-\!\left( \xi_{{\bf k}}\!-\!\xi_{{\bf p}}
\right)^{2} \right\}^{-1} .
\label{asym_J}
\end{eqnarray}
For the half-filled quantum dot ($\varepsilon_{d}\!=\!-U_{d}/2$) 
its value near the Fermi surface becomes negative
\begin{eqnarray}
J_{{\bf k}_F {\bf p}_F}(l \rightarrow \infty)
= \frac {- 4 U_{d} |V_{{\bf k}_{F}}|^2}{U_{d}^{2} + 
\left(2\Delta_{d}\right)^2}
\label{lowest}
\end{eqnarray}
and this fact is crucial for the Kondo effect.

\subsection{Inter-species pairing}

\begin{figure}
\epsfxsize=9.5cm{\epsffile{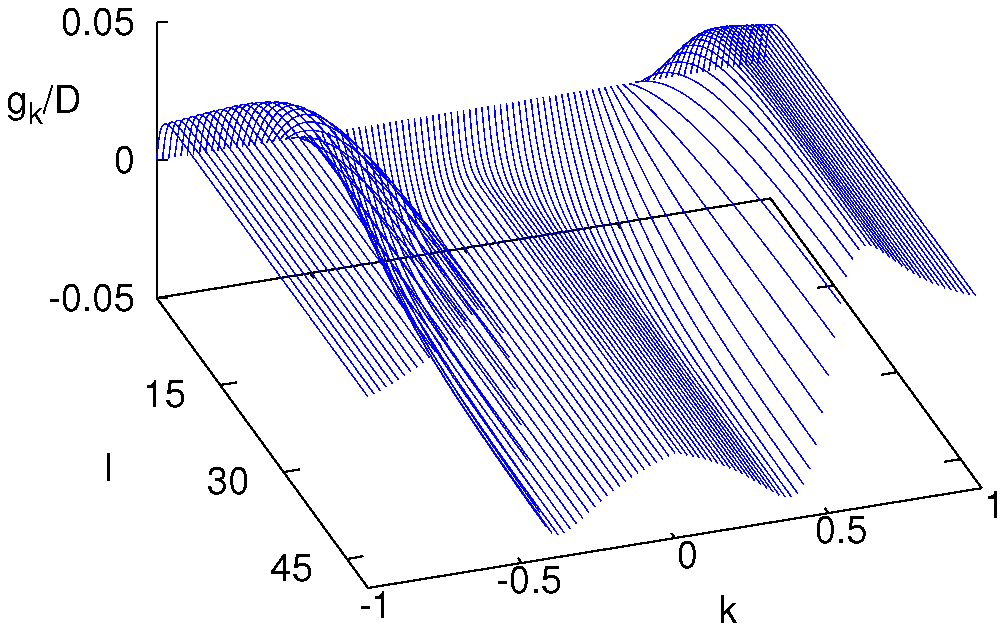}}
\caption{The inter-species pairing potential $g_{\bf k}^{\sigma}(l)$ 
obtained numerically for the half-filled quantum dot using 
the same set of model parameters as in the main text.}
\label{gk_flow}
\end{figure}
\begin{figure}
\epsfxsize=8cm{\epsffile{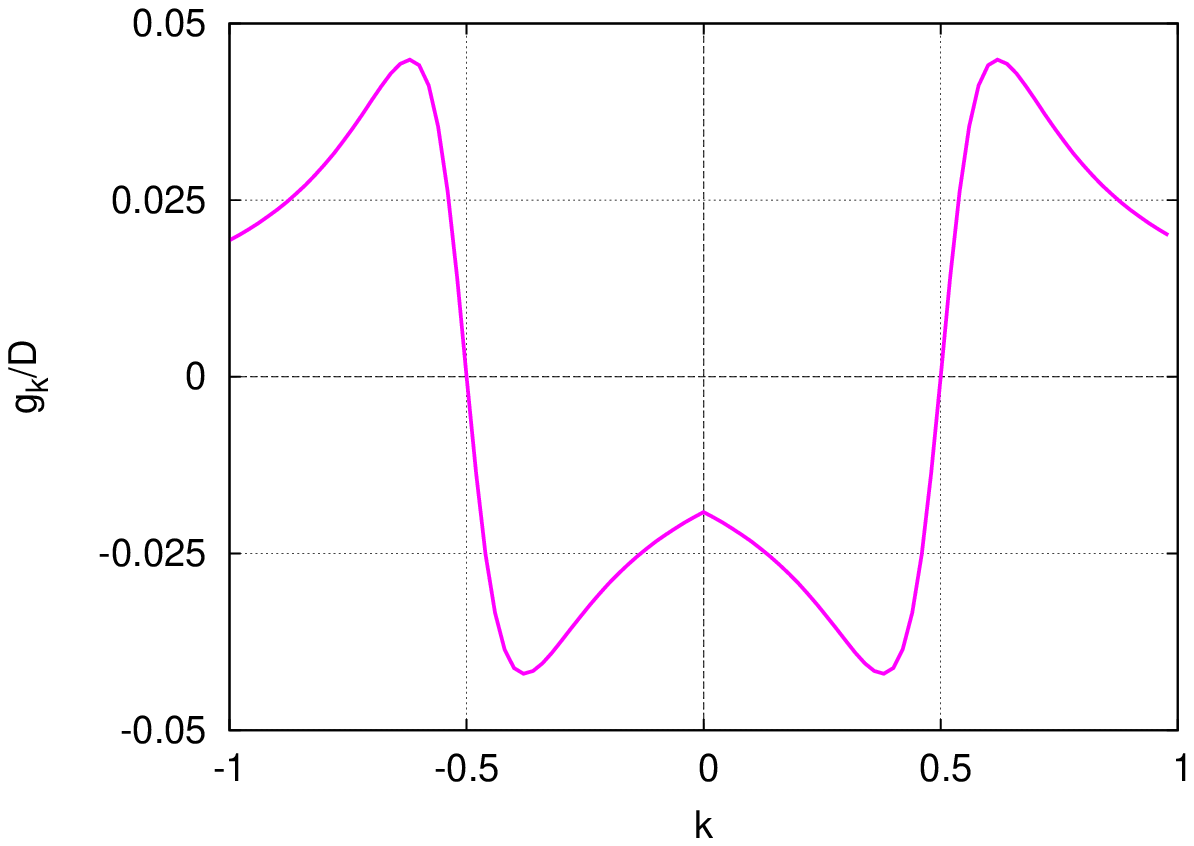}}
\caption{The asymptotic value $g_{\bf k}^{\sigma}(\infty)$ 
of the potential displayed in figure \ref{gk_flow}.}
\label{gk_infty}
\end{figure}
Besides the induced spin exchange interaction there can 
appear also other kinds of interactions. To give an example 
how such interactions can be studied in a systematic way  
we shall discuss here the exotic inter-species coupling
specific for the proximized quantum dot (\ref{proximized}). 
To take it into account we introduce the following 
$l$-dependent Hamiltonian 
\begin{eqnarray}
\hat{H}(l) &=& \sum_{{\bf k} \sigma} \xi_{\bf k}(l) 
\hat{c}_{{\bf k}\sigma}^{\dagger} \hat{c}_{{\bf k}\sigma}
+ \sum_{\sigma} \varepsilon_{d}(l) \hat{d}_{\sigma}^{\dagger} 
\hat{d}_{\sigma} + U_{d}(l){\hat n}_{d, \uparrow} 
{\hat n}_{d, \downarrow} \nonumber \\ 
&-&  \left( \Delta_{d}(l) \hat{d}_{\uparrow}^{\dagger} 
\hat{d}_{\downarrow}^{\dagger} + \mbox{\rm h.c.} \right)  
-  \sum_{{\bf k}, {\bf p}} 
J_{\bf k p}(l) \hat{\bf s}_{d} \cdot \hat{\bf S}_{{\bf k}{\bf p}}
\nonumber \\
 &-& \sum_{\bf k}  \left[ \left( g_{\bf k}^{\uparrow}(l)  
  \hat{c}_{{\bf k} \uparrow}^{\dagger} 
  \hat{d}_{\downarrow}^{\dagger}  + g_{\bf k}^{\downarrow}(l) 
 \hat{d}_{\uparrow}^{\dagger}\hat{c}_{\bf k \downarrow}^{\dagger}
 \right) +\mbox{\rm h.c.}\right ] 
\nonumber \\
&+& \sum_{{\bf k} \sigma} \left(  V_{\bf k}(l)
\hat{c}_{{\bf k}\sigma}^{\dagger} \hat{d}_{\sigma} +
\mbox{\rm h.c.} \right) 
\label{hamil_gk_flow}
\end{eqnarray}
with the initial condition $g^{\sigma}(0)=0$. Repeating the same procedure
as discussed in the previous section we obtain the additional flow equation
for inter-species coupling
\begin{eqnarray}
\frac{d g_{\bf k}^{\sigma}(l)}{d l} = \left [ U_{d}(l)\left 
( 2 \langle {\hat n}_{d,\sigma} \rangle - 1 \right )
+ 2\xi_{\bf k} \right ] \Delta_{d}(l) V_{\bf k}(l).
\label{inter}
\end{eqnarray}
For the half-filled quantum dot  $\langle {\hat n}_{d,\sigma} 
\rangle=\frac{1}{2}$ this equation (\ref{inter}) simplifies to
\begin{eqnarray}
\frac{d g_{\bf k}^{\sigma}(l)}{d l} = 2\xi_{\bf k} 
\Delta_{d}(l) V_{\bf k}(l).
\end{eqnarray}
Figure (\ref{gk_flow}) shows the coupling $g^{\sigma}_{\bf k}(l)$
obtained numerically for the half-filled quantum dot. The next plot 
(\ref{gk_infty}) illustrates the effective (asymptotic) value
$g^{\sigma}_{\bf k}(\infty)$. We clearly notice that the inter-species 
pairing $g_{\bf k}^{\sigma}$  changes the sign around $k_F$. This 
property indicates the resonant character of such exotic interactions.